\documentstyle[aps,prl,twocolumn]{revtex}
\def\beq{\begin{equation}}
\def\eeq{\end{equation}}
\def\bea{\begin{eqnarray}}
\def\eea{\end{eqnarray}}
\def\nn{\nonumber}
\def\sss{\scriptscriptstyle}

\def\bd{B_d^0}
\def\bdbar{{\overline{B_d^0}}}
\def\bs{B_s^0}
\def\bsbar{{\overline{B_s^0}}}
\def\bbar{{\overline{B^0}}}
\def\barp{{\raise.35ex\hbox
{${\sss (}$}}---{\raise.35ex\hbox{${\sss )}$}}}
\def\bdbarp{\hbox{$B_d$\kern-1.4em\raise1.4ex\hbox{\barp}}}
\def\bsbarp{\hbox{$B_s$\kern-1.4em\raise1.4ex\hbox{\barp}}}
\def\barpd{{\raise.35ex\hbox
{${\sss (}$}}--{\raise.35ex\hbox{${\sss )}$}}}
\def\dbarp{\hbox{$D^{*0}$\kern-1.6em\raise1.5ex\hbox{\barpd}}}
\def\kbarp{\hbox{$K^{*0}$\kern-1.6em\raise1.5ex\hbox{\barpd}}}
\def\ks{K_{\sss S}}

\def\roughly#1{\mathrel{\raise.3ex\hbox
{$#1$\kern-.75em\lower1ex\hbox{$\sim$}}}}

\def\adir00{{a_{\sss dir}^{00}}}

\def\B00{B^{00}}
\def\Bp0{B^{+0}}

\def\epjc#1#2#3{{\it Eur.\ Phys.\ J.}\ {\bf C#1}, #3 (19#2)}

\def\npb#1#2#3{{\it Nucl.\ Phys.} {\bf B#1}, #3 (19#2)}
\def\plb#1#2#3{{\it Phys.\ Lett.} {\bf #1B}, #3 (19#2)}
\def\prd#1#2#3{{\it Phys.\ Rev.} {\bf D#1}, #3 (19#2)}

\def\newprdtwo#1#2#3{{\it Phys.\ Rev.} {\bf D#1}: #3 (20#2)}

\def\prl#1#2#3{{\it Phys.\ Rev.\ Lett.} {\bf #1}, #3 (19#2)}
\def\zpc#1#2#3{{\it Zeit.\ Phys.} {\bf C#1}, #3 (19#2)}
\def\ijmp#1#2#3{{\it Int.\ J.\ Mod.\ Phys.} {\bf A#1}, #3 (19#2)}

\begin{document}
\draft 
\twocolumn[\hsize\textwidth\columnwidth\hsize\csname
  @twocolumnfalse\endcsname
\vspace{-0.5in}
\begin{flushright}
UdeM-GPP-TH-00-72 \\IMSc-00/05/17
\end{flushright}
\title{Extracting Weak Phase Information from $B \to V_1 V_2$ Decays}
\author{David London} 
\address { Laboratoire Ren\'e J.-A. L\'evesque, 
Universit\'e de Montr\'eal, C.P. 6128, succ. centre-ville, Montr\'eal, QC,
Canada H3C 3J7.\cite{e1}}
\author{Nita Sinha and Rahul Sinha}
\address {The Institute of Mathematical
Sciences, C. I. T Campus, Taramani, Chennai 600 113, India.\cite{e2}}  
\date{\today} 
\maketitle
\begin{abstract}
  We describe a new method for extracting weak, CP-violating phase
  information, with no hadronic uncertainties, from an angular
  analysis of $B \to V_1 V_2$ decays, where $V_1$ and $V_2$ are vector
  mesons. The quantity $\sin^2 (2\beta + \gamma)$ can be cleanly
  obtained from the study of decays such as $\bd(t) \to D^{*\pm}
  \rho^\mp$, $D^{*\pm} a_1^{\mp}$, $\dbarp~~\kbarp ~~$, etc.
  Similarly, one can use $\bs(t) \to D_s^{*\pm} K^{*\mp}$ to extract
  $\sin^2 \gamma$.  There are no penguin contributions to these
  decays. It is possible that $\sin^2 (2\beta + \gamma)$ will be the
  second function of CP phases, after $\sin 2\beta$, to be measured at
  $B$-factories.
\end{abstract}
\pacs{PACS number: }
] 

One of the most important open questions in particle physics is the
origin of CP violation. According to the standard model (SM), CP
violation is due to the presence of a nonzero complex phase in the
Cabibbo-Kobayashi-Maskawa (CKM) quark mixing matrix. This explanation
can be tested in the $B$ system. By measuring CP-violating rate
asymmetries in $B$ decays, one can extract $\alpha$, $\beta$ and
$\gamma$, the three interior angles of the unitarity triangle
\cite{CPreview}. The measured values of these angles may be consistent
with SM predictions, or they may indicate the presence of physics
beyond the SM. Hopefully, it is this latter scenario which will be
realized.

The reason that $B$ decays are such a useful tool is that the CP
angles can be obtained without hadronic uncertainties. The usual
technique is to consider a final state $f$ to which both $B^0$ and
$\bbar$ can decay. Because of $B^0$--$\bbar$ mixing, CP violation then
comes about due to an interference between the amplitudes $B^0 \to f$
and $B^0 \to \bbar \to f$. In the early days of the field, it was
thought that the CP angles could be easily measured in $\bd(t) \to
\pi^+ \pi^-$ ($\alpha$), $\bd(t) \to \Psi\ks$ ($\beta$), and $\bs(t)
\to \rho\ks$ ($\gamma$). However, it soon became clear that things
would not be so easy: the presence of penguin amplitudes
\cite{penguins} makes the extraction of $\alpha$ from $\bd(t) \to
\pi^+ \pi^-$ quite difficult, and completely spoil the measurement of
$\gamma$ in $\bs(t) \to \rho\ks$.  Even in the gold-plated mode
$\bd(t) \to \Psi\ks$, penguin contributions limit the precision with
which $\beta$ can be measured to about 2\%.  In part because of this,
a great deal of work was then done developing new methods to cleanly
obtain the CP angles from a wide variety of final states.

One class of final states that was considered consists of two vector
mesons, $V_1 V_2$. Because the final state does not have a
well-defined orbital angular momentum, $V_1 V_2$ cannot be a CP
eigenstate. This then implies that, even if both $B^0$ and $\bbar$ can
decay to the final state $V_1 V_2$, one cannot extract a CP phase
cleanly. However, this situation can be remedied with the help of an
angular analysis \cite{helicity}. By examining the decay products of
$V_1$ and $V_2$, one can measure the various helicity components of
the final state.  Since each helicity state corresponds to a state of
well-defined CP, an angular analysis allows one to use $B \to V_1 V_2$
decays to obtain one of the CP phases cleanly. Thus, for example, the
angle $\beta$ can be extracted from the decay $\bd(t) \to \Psi K^*$:
each helicity state of $\Psi K^*$ can be treated in the same way as
$\Psi \ks$.

In this paper, we show that the angular analysis is more powerful than
has previously been realized. Due to the interference between the
different helicity states, there are enough independent measurements
that one can obtain weak phase information from the decays of $B^0$
and $\bbar$ to any common final state $f$. Furthermore, contrary to
other methods, it is not necessary to measure the branching ratios of
both $B^0 \to f$ and $\bbar \to f$. This is important for final states
such as $D^{*\pm} \rho^{\mp}$, in which one of the two decay
amplitudes is considerably smaller than the other one.

We consider a final state $f$, consisting of two vector mesons, to
which both $B^0$ and $\bbar$ can decay. We assume further that only
one weak amplitude contributes to $B^0 \to f$ and $\bbar \to f$. We
write the helicity amplitudes as follows:
\begin{eqnarray}
\label{amp1}
A_\lambda \equiv Amp (B^0 \to f)_\lambda &=& 
       a_\lambda e^{i\delta_\lambda^a} e^{i\phi_a} ~, \\ 
A'_\lambda \equiv Amp (\bbar \to f)_\lambda &=& 
       b_\lambda e^{i\delta_\lambda^b} e^{i\phi_b} ~, \\ 
{\bar A}'_\lambda \equiv Amp (B^0 \to {\bar f})_\lambda &=& 
       b_\lambda e^{i\delta_\lambda^b} e^{-i\phi_b} ~, \\ 
{\bar A}_\lambda \equiv Amp (\bbar \to {\bar f})_\lambda &=& 
       a_\lambda e^{i\delta_\lambda^a} e^{-i\phi_a} ~,
\label{amp4}
\end{eqnarray}
where the helicity index $\lambda$ takes the values $\left\{
  0,\|,\perp \right\}$. In the above, $\phi_{a,b}$ and
$\delta^{a,b}_\lambda$ are the weak and strong phases, respectively.

Using CPT invariance, the total decay amplitudes can be written as
\begin{eqnarray}
{\cal A} = Amp (B^0\to f) = A_0 g_0 + A_\| g_\| + i \, A_\perp g_\perp~~ ,
\label{A}\\
{\bar{\cal A}} = Amp (\bbar\to {\bar f}) = 
     {\bar A}_0 g_0 + {\bar A}_\| g_\| - i \, {\bar A}_\perp g_\perp~~ ,
\label{Abar}\\
{\cal A}' = Amp (\bbar\to f) = A'_0 g_0 + A'_\| g_\| - i \, A'_\perp g_\perp~~ ,
\label{Aprime}\\
{\bar{\cal A}}' = Amp (B^0 \to {\bar f}) = 
   {\bar A}'_0 g_0 + {\bar A}'_\| g_\| + i \, {\bar A}'_\perp g_\perp~~ ,
\label{Aprimebar}
\end{eqnarray}
where the $g_\lambda$ are the coefficients of the helicity amplitudes
written in the linear polarization basis. The $g_\lambda$ depend only
on the angles describing the kinematics \cite{flambda,flamdun}.

With the above equations, the time-dependent decay rate for $B^0(t)
\to f$ can be written as
\bea
\Gamma(B^0(t) \to f)= e^{-\Gamma t} \sum_{\lambda\leq\sigma}&&
\Bigl(\Lambda_{\lambda\sigma} + \Sigma_{\lambda\sigma}\cos(\Delta M t)\nn\\&&
- \rho_{\lambda\sigma}\sin(\Delta M t)\Bigr) g_\lambda g_\sigma ~.
\eea
Thus, by performing a time-dependent study and angular analysis of the
decay $B^0(t)\to\! f$, one can measure the observables
$\Lambda_{\lambda\sigma}$, $\Sigma_{\lambda\sigma}$ and
$\rho_{\lambda\sigma}$. In terms of the helicity amplitudes
$A_0,A_\|,A_\perp$, these can be expressed as follows:
%
\bea
&\Lambda_{\lambda\lambda}=\displaystyle
\frac{|A_\lambda|^2+|A'_\lambda|^2}{2},&
\Sigma_{\lambda\lambda}=\displaystyle
\frac{|A_\lambda|^2-|A'_\lambda|^2}{2},\nn \\[1.5ex]
&\Lambda_{\perp i}= -\!{\rm Im}({ A}_\perp { A}_i^* \!-\! A'_\perp {A'_i}^* ),
&\Lambda_{\| 0}= {\rm Re}(A_\| A_0^*\! +\! A'_\| {A'_0}^* ),
\nn \\[1.5ex]
&\Sigma_{\perp i}= -\!{\rm Im}(A_\perp A_i^*\! +\! A'_\perp {A'_i}^* ),
&\Sigma_{\| 0}= {\rm Re}(A_\| A_0^*\!-\! A'_\| {A'_0}^* ),\nn\\[1.5ex]
&\rho_{\perp i}\!=\!-\!{\rm Re}\!\Bigl(\!\frac{q}{p}
\![A_\perp^*  A'_i\! +\! A_i^* A'_\perp\!]\!\Bigr),
&\rho_{\perp \perp}\!=\! -\! {\rm Im}\Bigl(\frac{q}{p}\,
A_\perp^* A'_\perp\Bigr),\nn\\[1.5ex]
&\rho_{\| 0}\!=\!{\rm Im}\!\Bigl(\frac{q}{p}\!
[A_\|^* A'_0\! + \!A_0^* A'_\|\!]\!\Bigr),
&\rho_{ii}\!=\!{\rm Im}\!\Bigl(\frac{q}{p} A_i^* A'_i\Bigr),
\label{defs}
\eea
where $i=\{0,\|\}$. In the above, $q/p = \exp({-2\,i\phi_{\sss M}})$,
where $\phi_{\sss M}$ is the weak phase present in $B^0$--$\bbar$
mixing.

Similarly, the decay rate for $B^0(t) \to {\bar f}$ is given by
%
\bea
\Gamma(B^0(t)\to {\bar f}) = e^{-\Gamma t} \sum_{\lambda\leq\sigma}&&
\Bigl(
{\bar\Lambda}_{\lambda\sigma} + {\bar\Sigma}_{\lambda\sigma}\cos(\Delta M t) \nn\\&&
- {\bar\rho}_{\lambda\sigma}\sin(\Delta M t)
\Bigr)  g_\lambda  g_\sigma ~.
\eea
The expressions for the observables ${\bar\Lambda}_{\lambda\sigma}$,
${\bar\Sigma}_{\lambda\sigma}$ and ${\bar\rho}_{\lambda\sigma}$ are
similar to those given in Eq.~(\ref{defs}), with the replacements
$A_\lambda \to {\bar A}'_\lambda$ and $A'_\lambda \to {\bar
  A}_\lambda$.

With the above expressions for the various amplitudes, we now show how
to extract weak phase information using the above measurements. First,
we note that
\beq
\Lambda_{\lambda\lambda}={\bar\Lambda}_{\lambda\lambda}=
\frac{(a_\lambda^2+b_\lambda^2)}{2},
\Sigma_{\lambda\lambda}=-{\bar\Sigma}_{\lambda\lambda}=
\frac{(a_\lambda^2-b_\lambda^2)}{2}.
\label{LamSig_eq}
\eeq
Thus, one can determine the magnitudes of the amplitudes appearing in
Eqs.~(\ref{amp1})--(\ref{amp4}), $a_\lambda^2$ and $b_\lambda^2$.
However, we must stress that, in fact, knowledge of $b_\lambda^2$ will
not be necessary within our method. This is important since some final
states have $b_\lambda \ll a_\lambda$, and so the determination of
$b_\lambda^2$ would be very difficult.

Next, we have
\bea
&&\Lambda_{\perp i}\! =\! -{\bar\Lambda}_{\perp i}\! =\! b_{\perp} b_i
\sin(\delta_{\perp}\!-\!\delta_i\!+\!\Delta_i)-a_\perp a_i\sin(\Delta_i), \nn \\
&&\Sigma_{\perp i}\! =\! {\bar\Sigma}_{\perp i}\! =\! -b_\perp b_i
\sin(\delta_\perp\!-\!\delta_i\!+\!\Delta_i)-a_\perp a_i\sin(\Delta_i),
\label{LS}
\eea
where $\Delta_i \equiv \delta_\perp^a-\delta_i^a$ and $\delta_\lambda
\equiv \delta_\lambda^b-\delta_\lambda^a$. Using Eq.~(\ref{LS}) one
can solve for $a_\perp a_i\sin\Delta_i$. We will see that this is the
only combination needed to cleanly extract weak phase information.

The coefficients of the $\sin(\Delta m t)$ term, which can be
obtained in a time-dependent study, can be written as
\beq
\rho_{\lambda\lambda}\! =\! \pm a_\lambda b_\lambda \sin(\phi\!+\!\delta_\lambda),
{\bar\rho}_{\lambda\lambda}\!=\!\pm a_\lambda b_\lambda \sin(\phi\!-\!\delta_\lambda),
\label{rho_eq}
\eeq
where the sign on the right hand side is positive for $\lambda=\|,0$
and negative for $\lambda=\perp$. In the above, we have defined the CP
phase $\phi \equiv -2\phi_{\sss M} + \phi_b - \phi_a$. These
quantities can be used to determine
\beq
2 b_\lambda\cos\delta_\lambda\! =\! 
\pm\frac{\rho_{\lambda\lambda}\!+\!{\bar\rho}_{\lambda\lambda}}{a_\lambda \sin\phi},
2 b_\lambda\sin\delta_\lambda \!=\!
\pm\frac{\rho_{\lambda\lambda}\!-\!{\bar\rho}_{\lambda\lambda}}{a_\lambda \cos\phi}.
\label{delta}
\eeq

Similarly, the terms involving interference of different helicities
are given as
\begin{eqnarray}
\rho_{\perp i}\! &=&\! -a_\perp b_i \cos(\phi\!+\!\delta_i\!-\!\Delta_i)\!-\!
a_i b_\perp \cos(\phi\!+\!\delta_\perp\!+\!\Delta_i) , \nn\\
{\bar\rho}_{\perp i}\! &=&\! -a_\perp b_i  \cos(\phi\!-\!\delta_i\!+\!\Delta_i) \!-\! 
a_i b_\perp \cos(\phi\!-\!\delta_\perp\!-\!\Delta_i) .
\label{rhocombs}
\end{eqnarray}
%
%

Putting all the above information together, we are now in a position
to extract the weak phase $\phi$. Using Eq.~(\ref{delta}), the
expressions in Eq.~(\ref{rhocombs}) can be used to yield
\begin{eqnarray}
&&\rho_{\perp i}\!+\!{\bar\rho}_{\perp i}=
-\cot\phi\,{a_i a_\perp}\cos\Delta_i\Bigg[\frac{\rho_{i
i}+{\bar\rho}_{i i}}{a_i^2}- \frac{\rho_{\perp
\perp}+{\bar\rho}_{\perp \perp}}{a_\perp^2}\Bigg] \nn \\ &&
~~~~~~~~~~~~-{a_i a_\perp}\sin\Delta_i\Bigg[\frac{\rho_{i i}-{\bar\rho}_{i
i}}{a_i^2}+ \frac{\rho_{\perp \perp}-{\bar\rho}_{
\perp \perp}}{a_\perp^2}\Bigg],\\ 
&&\rho_{\perp i}\!-\!{\bar\rho}_{\perp i}=
\tan\phi\,{a_i a_\perp}\cos\Delta_i\Bigg[\frac{\rho_{i
i}-{\bar\rho}_{i i}}{a_i^2}- \frac{\rho_{\perp
\perp}-{\bar\rho}_{\perp \perp}}{a_\perp^2}\Bigg] \nn \\ &&
~~~~~~~~~~~~-{a_i a_\perp}\sin\Delta_i\Bigg[\frac{\rho_{i i}+{\bar\rho}_{i
i}}{a_i^2}+ \frac{\rho_{\perp \perp}+{\bar\rho}_{
\perp \perp}}{a_\perp^2}\Bigg].
\label{sol}
\end{eqnarray}
Now, we already know most of the quantities in the above two
equations: (i) $\rho_{\lambda\sigma}$ and ${\bar\rho}_{\lambda\sigma}$
are measured quantities, (ii) the $a_\lambda^2$ are determined from
the relations in Eq.~(\ref{LamSig_eq}), and (iii) ${a_i
  a_\perp}\sin\Delta_i$ is obtained from Eq.~(\ref{LS}). Thus, the
above two equations involve only two unknown quantities --- $\tan\phi$
and ${a_i a_\perp}\cos\Delta_i$ --- and can easily be solved (up to a
sign ambiguity in each of these quantities). In this way $\tan^2\phi$
(or, equivalently, $\sin^2 \phi$) can be obtained from the angular
analysis.

Note that this method relies on the measurement of the interference
terms between different helicities. However, we do not actually
require that all three helicity components of the amplitude be used.
In fact, one can use observables involving any two of largest helicity
amplitudes. In the above description, one could have chosen `$\|\,0$'
instead of `$\perp\!\|$' or `$\perp\! 0$'.

We now turn to specific applications of this method. Consider first
the situation in which the final state is a CP eigenstate, $f = \pm
{\bar f}$. In this case, the parameters of
Eqs.~(\ref{amp1})--(\ref{amp4}) satisfy $a_\lambda = b_\lambda$,
$\delta_\lambda^a = \delta_\lambda^b$ (which implies that
$\delta_\lambda = 0$), and $\phi_a = -\phi_b$ (so that $\phi \equiv
-2\phi_{\sss M} + 2 \phi_b$). As described above, $a_\lambda^2$ can be
obtained from Eq.~\ref{LamSig_eq}. But now the measurement of
$\rho_{\lambda\lambda}$ [Eq.~(\ref{rho_eq})] directly yields $\sin
\phi$. In fact, this is the conventional way of using the angular
analysis to measure the weak phases: each helicity state separately
gives clean CP-phase information. Thus, when $f$ is a CP eigenstate,
nothing is gained by including the interference terms.

Of course, in general, final states that are CP -- eigenstates will all
receive penguin contributions at some level.
 Thus, these states violate our assumption that only one
weak amplitude contributes to $B^0 \to f$ and $\bbar \to f$. The only
quark-level decays which do not receive penguin contributions are
${\bar b} \to {\bar c} u {\bar d} ,~ {\bar u} c {\bar d}$, as well as
their Cabibbo-suppressed counterparts, ${\bar b} \to {\bar c} u {\bar
  s} ,~ {\bar u} c {\bar s}$. These are, in fact, the types of decays
for which our method is most useful, and we will give meson-level
examples of each of these below.

Consider first the decays $\bd/\bdbar \to D^{*-}\rho^+, D^{*+}\rho^-$
(which correspond to ${\bar b} \to {\bar c} u {\bar d},~ {\bar u} c
{\bar d}$ at the quark level). In this case we have $\phi_{\sss M} =
\beta$, $\phi_a = 0$ and $\phi_b = -\gamma$, so that $\phi = - 2 \beta
- \gamma$. The method described above allows one to extract $\sin^2
(2\beta + \gamma)$ from an angular analysis of the final state
$D^{*\pm} \rho^\mp$.

In Ref.~\cite{BDpi}, Dunietz pointed out that $\sin^2 (2\beta +
\gamma)$ could, in principal, be obtained from measurements of $\bd(t)
\to D^\mp \pi^\pm$. He used the method of Ref.~\cite{ADKL}, which
requires the accurate measurement of the quantity $\Gamma(\bdbar \to
D^- \pi^+) / \Gamma(\bd \to D^- \pi^+)$. This ratio is essentially
$|V_{ub} V_{cd}^* / V_{cb}^* V_{ud}|^2 \simeq 4 \times 10^{-4}$.
Obviously, it will be very difficult to measure this tiny quantity
with any precision, which creates a serious barrier to carrying out
Dunietz's method in practice.

On the other hand, our method does not suffer from this problem. In
our notation [Eqs.~(\ref{amp1})--(\ref{amp4})], the rate
$\Gamma(\bdbar \to D^{*-} \rho^+)$ is proportional to $b_\lambda^2$.
However, as we have already emphasized in the discussion following
Eq.~(\ref{LamSig_eq}), a determination of this quantity is not needed
to extract $\sin^2 (2\beta + \gamma)$ using the angular analysis: {\it
none of the observables or combinations required for the analysis are
proportional to $b_\lambda^2$}. Thus, we avoid the practical problems
present in Dunietz's method.

One disadvantage of the final states $D^{*\pm} \rho^\mp$ is that the
two decay amplitudes are very different in size (hence the small value
of $b_\lambda$). This results in a very small CP-violating asymmetry
whose size is approximately $|V_{ub} V_{cd}^* / V_{cb}^* V_{ud}|
\approx 2\%$. Since the total number of $B$'s required to make the
measurement is inversely proportional to the square of the asymmetry
$A_f$, $N_{\sss B} \propto 1/(BR(\bd\to f) \, A_f^2)$,
this is a potential problem, even though the branching ratio for the
decay $\bd \to D^{*-}\rho^+$ is quite large, roughly 1\%.

One can avoid the problem of a small asymmetry by instead using the
Cabibbo-suppressed decays $\bd \to {\bar D}^{*0} K^{*0}, D^{*0}
K^{*0}$ and $\bdbar \to D^{*0} {\bar K}^{*0}, {\bar D}^{*0} {\bar
K}^{*0}$ (corresponding to the quark-level decays ${\bar b} \to {\bar
c} u {\bar s} ,~ {\bar u} c {\bar s}$) \cite{BDK}. (Here it is
assumed that both $K^{*0}$ and ${\bar K}^{*0}$ decay to the same state
$\ks \pi^0$.) In this case the two amplitudes are much more equal in
size, leading to a large asymmetry of about $|V_{ub} V_{cs}^* /
V_{cb}^* V_{us}| \approx 40\%$. The disadvantage, of course, is that
the branching ratios for such Cabibbo-suppressed decays are much
smaller than those for $\bd/\bdbar \to D^{*\pm}\rho^\mp$. We estimate
that $B(\bd \to {\bar D}^{*0} K^{*0}) \approx \lambda^2 B(\bd\to\Psi
K^{*0}) = 7 \times 10^{-5}$, which, when combined with $B(K^{*0} \to
\ks\pi^0) = 1/3$, yields a net branching ratio of about $2 \times
10^{-5}$. Even though this branching ratio is quite a bit smaller than
that for $\bd \to D^{*-}\rho^+$, the much larger asymmetry makes up
for it. 
We see that the measurement of
$\sin^2(2\beta+\gamma)$ using $\bd(t) \to \dbarp~\kbarp~$ requires
roughly the same number of $B$'s as if $\bd(t) \to D^{*\pm}\rho^\mp$
were used.

Of course, this leads to an important question: just how many $B$'s
are needed for such a measurement? The CLEO collaboration has already
performed an angular analysis of $\bd \to D^{*-}\rho^+$ with a sample
of $197\pm15$ events \cite{CLEOII}, and has been able to measure some of the
interference terms. (Of course, since they do not have an asymmetric
collider, it is not possible for them to measure the $\sin(\Delta M
t)$ terms.) In addition, in our method it is necessary to tag the
decaying $\bd$/$\bdbar$. Taking the tagging efficiency to be about
30\% \cite{CPreview}, and using the above values for the branching
ratios and asymmetries, we estimate the total number of $B$'s required
to measure $\sin^2(2\beta+\gamma)$ using our method to be roughly
$10^8$. This number may be reduced if it is possible to combine the
various final states ($D^{*\pm} \rho^\mp$, $D^{*\pm} a_1^{\mp}$,
$\dbarp~~\kbarp ~~$, etc.). We therefore conclude that this
measurement will probably be possible at a first-generation
$B$-factory, though it may take several years of data accumulation.

In fact, the extraction of $\sin^2(2\beta+\gamma)$ may well turn out
to be the second clean measurement to be made at $B$-factories ($\sin
2\beta$ will clearly be measured first via $\bd(t) \to \Psi\ks$). As
discussed above, the angle $\alpha$ cannot be obtained cleanly from
$\bd(t) \to \pi^+ \pi^-$ due to the presence of penguin contributions.
This difficulty can be resolved with the aid of an isospin analysis
\cite{isospin}, but this technique requires measuring the branching
ratio for $\bd\to\pi^0\pi^0$, which may be quite small. It is also
possible to extract $\alpha$ with no hadronic uncertainties using a
Dalitz-plot analysis of $\bd(t)\to\pi^+\pi^-\pi^0$ decays
\cite{Dalitz}. Here the idea is to isolate the resonant contributions
from intermediate $\rho\pi$ states, to which certain isospin relations
apply. However, one has to be sure that the non-resonant contributions
are well-understood, which requires some theoretical input. In any
case, it is estimated that this measurement will take roughly six
years to complete. As for the angle $\gamma$, the original suggestion
for measuring it cleanly involved the decays $B^\pm \to D^0 K^\pm ,~
{\overline{D^0}} K^\pm ,~ D^0_{\sss CP} K^\pm$ \cite{GroWyler}.
However, it was subsequently shown that this type of analysis runs
into problems because it is virtually impossible to tag the flavor of
the final-state $D$-meson \cite{ADS}, and so one cannot distinguish
$B^\pm \to D^0 K^\pm$ from $B^\pm\to {\overline{D^0}} K^\pm$ decays.
One can still obtain $\gamma$ cleanly by studying decays such as $B^+
\to (K^+ \pi^-)_{\sss D} K^+$ and $B^+ \to (K^+ \rho^-)_{\sss D} K^+$,
along with their CP-conjugates, but this requires many more $B$'s, so
that it is unlikely such measurements can be carried out in the first
round of $B$-factory experiments. Finally, there has been much work
recently looking at the possibilities for extracting $\gamma$ from
$B\to\pi K$ decays \cite{BpiKreview}. However, all of these methods
use flavor $SU(3)$ symmetry, and so rely heavily on theoretical
input. In view of all of this, it is thus quite conceivable that the
second clean extraction of CP phases at $B$ factories will be the
measurement of $\sin^2(2\beta+\gamma)$ using the method described in
this paper.

Note that the measurement of $\sin^2(2\beta+\gamma)$ may turn out to
be very useful in looking for physics beyond the SM. If new physics is
present, it will affect the CP asymmetries principally through its
contributions to $B^0$--$\bbar$ mixing \cite{NPBmixing}.  The most
straightforward way of searching for this new physics is to consider
two distinct decay modes which, in the SM, probe the same CP angle. A
discrepancy between the two values would be clear evidence of physics
beyond the SM. For example, the angle $\gamma$ can be measured using
rate asymmetries in $B^\pm$ decays as described above ($B^\pm \to D
K^\pm$ \cite{GroWyler,ADS}), or in $\bs$/$\bsbar$ decays ($\bs(t) \to
D_s^\pm K^\mp$ \cite{ADK} or $\bs(t) \to D_s^{*\pm} K^{*\mp}$ [see
below]). If there is new physics in $\bs$--$\bsbar$ mixing, with new
phases, one will obtain different values of $\gamma$ from these two
systems. Unfortunately, as argued above, it will be difficult to use
$B^\pm$ decays to obtain $\gamma$, at least in the short term, so that
we will not have two independent values of $\gamma$ to compare.
However, this is where the measurement of $\sin^2(2\beta+\gamma)$ will
be useful: using the value of $2\beta$ as measured in $\bd(t) \to
\Psi\ks$, one can obtain $\gamma$, up to discrete ambiguities. If
none of these values of $\gamma$ coincide with those given by the
measurement of $\sin^2 \gamma$ in the $B_s$ system, this will be a
clear signal of new physics.

Finally, one can also consider $\bs$ and $\bsbar$ decays corresponding
to the quark-level decays ${\bar b} \to {\bar c} u {\bar d} ,~ {\bar
  u} c {\bar d}$, or ${\bar b} \to {\bar c} u {\bar s} ,~ {\bar u} c
{\bar s}$. The most promising processes are the Cabibbo-suppressed
decay modes $\bs/\bsbar \to D_s^{*\pm} K^{*\mp}$. Here $\phi_{\sss M}
= 0$, so that the quantity $\sin^2 \gamma$ can be extracted from the
angular analysis of $\bs(t)\to D_s^{*\pm} K^{*\mp}$. This is therefore
a new method of obtaining the CP phase $\gamma$. Note that $\sin^2
\gamma$ can also be obtained from a measurement of $\bs(t)\to D_s^\pm
K^\mp$ using a different method \cite{ADK}. The advantage of our
method is that the branching ratios are likely to be larger. On the
other hand, one must also perform an angular analysis, which is likely
to require more $B$'s. We therefore conclude that the two methods will
probably be of equal difficulty experimentally. Thus, this gives two
independent ways of extracting $\sin^2 \gamma$ from similar final
states.

In summary, we have presented a new method of using the angular
analysis of $B\to V_1 V_2$ decays to extract weak, CP-violating phases
with no hadronic uncertainties. Its most useful application involves
the quark-level decays ${\bar b} \to {\bar c} u {\bar d} ,~ {\bar u} c
{\bar d}$, and ${\bar b} \to {\bar c} u {\bar s} ,~ {\bar u} c {\bar
  s}$. We have shown that the quantity $\sin^2 (2\beta + \gamma)$ can
be cleanly obtained from the study of the decays $\bd(t) \to D^{*\pm}
\rho^\mp$, $D^{*\pm} a_1^{\mp}$, $\dbarp~~\kbarp ~~$, etc.  Similarly,
$\sin^2 \gamma$ can be extracted from $\bs(t) \to D_s^{*\pm}
K^{*\mp}$. In all of these cases, there are no penguin contributions
to the decays. Finally, we have argued that, due to difficulties with
other methods of measuring CP phases, $\sin^2 (2\beta + \gamma)$ may
well be the second clean measurement, after $\sin 2\beta$, which will
be made at $B$-factories.

N.S. and R.S. thank D.L. for the hospitality of the Universit\'e de
Montr\'eal, where part of this work was done. The work of D.L. was
financially supported by NSERC of Canada.

\end{document}